\documentclass[jrfm,article,accept,pdftex,moreauthors]{Definitions/mdpi} 
\graphicspath{ {./figures/} }

\usepackage[labelformat=simple]{subcaption}

\DeclareCaptionLabelFormat{subcaptionlabel}{\normalfont(\textbf{#2}\normalfont)}
\usepackage{threeparttable}
\usepackage{amsfonts}
\usepackage{rotating}
\usepackage{appendix}
\usepackage[acronym]{glossaries}
\newacronym{rl}{RL}{Reinforcement Learning}
\newacronym{mdp}{MDP}{Markov Decision Process}
\newacronym{dqn}{DQN}{Deep Q-Learning}
\newacronym{pg}{PG}{Policy Gradients}
\newacronym{a2c}{A2C}{Advantage Actor-Critic}
\newacronym{ppo}{PPO}{Proximal Policy Optimization}
\newacronym{ddpg}{DDPG}{Deep Deterministic Policy Gradient}
\newacronym{sac}{SAC}{Soft Actor Critic}
\newacronym{td3}{TD3}{Twin Delayed DDPG}
\newacronym{optics}{OPTICS}{Ordering Points To Identify The Clustering Structure}
\newacronym{credit}{CREDIT}{reCurrent Reinforcement lEarning methoD for paIrs Trading}
\newacronym{ssd}{SSD}{Sum of Squared Deviation}
\newacronym{cagr}{CAGR}{Compound Annual Growth Rate}
\newacronym{pnl}{P\&L}{Profit and Loss}
\newacronym{ai}{AI}{Artificial Intelligence}
\newacronym{adf}{ADF}{Augmented Dickey-Fuller}

\firstpage{1} 
\pubvolume{17}
\issuenum{12}
\articlenumber{555}
\pubyear{2024}
\copyrightyear{2024}
\externaleditor{Academic Editor: Xianrong (Shawn)~Zheng}
\datereceived{8 November 2024} 
\daterevised{28 November 2024} 
\dateaccepted{6 December 2024} 
\datepublished{11 December 2024} 
\hreflink{\textls[-25]{https://doi.org/10.3390/jrfm17120555}}

\Title{Reinforcement Learning Pair Trading: A Dynamic Scaling~Approach}

\TitleCitation{Reinforcement Learning Pair Trading: A Dynamic Scaling Approach}

\Author{Hongshen Yang * and Avinash Malik}

\AuthorCitation{Yang, Hongshen, and Avinash Malik}

\address[1]{Department of ECSE, The University of Auckland, Auckland 1010
, New Zealand; avinash.malik@auckland.ac.nz}

\corres{\hangafter=1 \hangindent=1.05em \hspace{-0.82em} Correspondence: hyan212@aucklanduni.ac.nz}

\abstract{Cryptocurrency is a cryptography-based digital asset with extremely
  volatile prices. Around USD 70 billion worth of cryptocurrency is
  traded daily on exchanges. Trading cryptocurrency is difficult due to
  the inherent volatility of the crypto market. This study investigates whether Reinforcement Learning (RL) can enhance decision-making in cryptocurrency algorithmic trading compared to traditional methods. In order to address this question, we combined reinforcement learning with a statistical arbitrage trading technique, pair trading, which exploits the price difference between statistically correlated assets. We constructed RL environments and trained RL agents to determine when and how to trade pairs of cryptocurrencies. We developed new reward
  shaping and observation/action spaces for reinforcement learning. We
  performed experiments with the developed reinforcement learner on pairs of BTC-GBP and BTC-EUR data separated by 1 min intervals
  ($n$~=~263,520). The traditional non-RL pair trading technique achieved an
  annualized profit of 8.33\%, while the proposed RL-based pair trading
  technique achieved annualized profits from 9.94\% to 31.53\%, depending
  upon the RL learner. Our results show that RL can significantly outperform manual and traditional pair trading techniques when applied
  to volatile markets such as~cryptocurrencies.}


\begin{document}

\section{Introduction\label{sec:introduction}}

Arbitrage is a subdomain of financial trading that profits from price
discrepancies in different markets \citep{dybvig_arbitrage_1989}. Pair
trading is one of the well-known arbitrage trading methods in financial
markets. Arbitrageurs identify two highly correlated assets to form a
pair. When a price discrepancy happens, they buy the underpriced asset
and sell the overpriced correlated asset to profit from the mean
reversion of the prices. With the rise of high-frequency trading, the ability to conduct fast and accurate analyses has become critical. Arbitrage requires practitioners to constantly
analyze the market conditions at the fastest speed possible, as
arbitrageurs must compete for transitory
opportunities~\mbox{\citep{brogaard_high-frequency_2014}}. Therefore, we explore how \acrfull{ai} can enhance the process of pair trading, focusing on the speed and adaptability of \mbox{decision-making.}

\acrfull{rl} is a captivating domain of \acrshort{ai}. The idea of
\acrshort{rl} is to let the agent(s) learn to interact with an
environment. The agent should learn from the environment's responses to
optimize its behavior \citep{sutton_reinforcement_2018}. If we view the
financial market from the perspective of the \acrshort{rl} environment,
actions in the financial market are investment decisions. By allowing agents to adapt dynamically to market conditions, \acrshort{rl} has the potential to overcome the limitations of static, rule-based strategies in volatile and complex financial environments. For gaining profits, arbitrageurs are incentivized to train agents to produce
lucrative investment decisions, and \acrshort{rl} facilitates agents'
learning process from the profit/loss of the market.

The combination of \acrshort{rl} and various financial trading techniques is still evolving rapidly. There has been some work in \acrshort{rl} infrastructural construction~\mbox{(\citeauthor{liu_finrl_2021}} \citeyear{liu_finrl_2021}, \citeyear{liu_finrl_2022}, \citeyear{liu_finrl-meta_2022}) and some experiments in profitable \acrshort{rl} agent training \citep{meng_reinforcement_2019, zhang_deep_2020, pricope_deep_2021}. Trading actions in traditional pair trading follow static rules.~In reality, the complexity of financial markets should allow more flexibility in the decision-making process.~An experienced trader might analyze market conditions to make informed decisions. However, it is not feasible to output efficient decisions at short, intermittent intervals 24/7. \acrshort{rl} algorithms enable a fast-track decision-making process for analyzing trading signals and generating trading actions.

Designing a high-frequency trading system based on \acrshort{rl} requires addressing critical challenges. The first challenge is how to construct an RL environment that accommodates RL agents for arbitrage. The second challenge involves identifying compatible instruments with historical correlations to form profitable pairs. The third challenge concerns timing. Instead of blindly following preset rules, the system requires flexibility in choosing investment timings to achieve greater profitability. The final challenge involves investment quantity. Since investment opportunities vary in quality, a critical consideration is whether RL agents can replicate decision-making capabilities comparable to the scrutiny applied by experienced traders.

This paper investigates key questions centered around the application of Reinforcement Learning (RL) in pair trading. To address the fast decision-making requirements in a high-frequency trading environment, we constructed a tailored RL environment for pair trading and fine-tuned reward shaping to encourage the agent to make profitable decisions. The contributions of this work are as follows:
(1) the construction of an RL environment specifically designed for quantity-varying pair trading;
(2) the proposal of a novel pair trading method that incorporates adaptive investment quantities to capture opportunities in highly volatile markets;
(3) the use of a grid search technique to fine-tune hyperparameters for enhanced profitability;
(4) the introduction of an RL component for market analysis and decision-making in pair trading, along with a novel RL model optimized for investment quantity decisions.

The structure of the paper is arranged as follows: the background and
related work are introduced in Sections~\ref{sec:background}
and~\ref{sec:relatedwork}. The methodology is presented in
Section~\ref{sec:methodology}. Experiments and results are included in
Section~\ref{sec:experiments}. A discussion of the results and conclusions is provided in~Section~\ref{sec:conclusion}.

\section{Background\label{sec:background}}

First, we define the basic terms of financial trading. A \textit{long
} position is created when an investor uses cash to buy an asset, and a
\textit{short
} position is created when an investor sells a borrowed
asset. The portfolio is the investor's total holding, including
long/short position and cash. \textit{Transaction cost
} is a percentage
fee payable to the broker for any long/short actions. Finally,
\textit{risk 
} is defined as the volatility of the portfolio.

\subsection{Traditional Pair Trading \label{sec:tradpairtrade}}

Classical pair trading consists of two distinct components known as
\textit{legs
}. A \textit{leg 
} represents one side of a trade in a
multi-contract trading strategy. Under the definition of pair trading,
``longing the first asset and shorting the second asset'' is called a
\textit{long leg
}, and ``shorting the first asset and longing the second
asset'' is called a \textit{short leg
}. The two assets are always bought
and sold in opposite directions in pair trading. Therefore, the overall
pair trading strategy is considered to be \textit{market neutral
}
because the profits from the \textit{long 
} position and the
\textit{short 
} are offset by the direction of the overall market.
\citeauthor{gatev_pairs_2006}'s~(\citeyear{gatev_pairs_2006}) work is the
most cited traditional pair trading method. It follows the \textit{OODA}
(observe, orient, decide, and act) loop \citep{fadok_air_1995}. Before
entering the market, the first step is to choose the proper assets in a
pair. The \acrfull{ssd} is the measurement calculated from the prices of assets
$i$ and $j$. Through exhaustive searching in a formation period $T$, the
assets with the smallest \acrshort{ssd} are bound as a pair
Equation~(\ref{equ:ssd}).

\begin{equation}
    \label{equ:ssd}
    SSD_{p_i,p_j}=\sum_{t=1}^{T}(p_i-p_j)^2.
\end{equation}

\begin{itemize}
\item \textbf{Observe
} is the process of market analysis. The price of
  assets in pairs is collected and processed. The price difference
  $(p_i-p_j)$ is called spread $S$. The arbitrageurs \textit{observe
}
  the current positions and spread of the current market.
\item \textbf{Orient 
} is the process exploring what could be done. Three possible
  actions for pair trading are long leg, short leg, and close position, as
  defined above.
\item \textbf{Decide 
} what action to take. Position opening triggers
  when the price difference deviates too much. This is indicated by the
  spread movement beyond an open threshold. Position closing happens
  when the spread reverts back to some closing threshold.
  \citeauthor{gatev_pairs_2006} (\citeyear{gatev_pairs_2006}) adopted
  two times the standard deviation of the spread as the opening
  threshold and the price crossing as the closing threshold. In
  practice, the threshold varies according to the characteristics of the
  financial instrument.
  \item \textbf{Act 
} once the decision is made. The long leg orders us to buy asset $i$ and sell asset $j$. The short leg orders us to sell asset $i$ and buy asset $j$. Closing a position means clearing all the active positions to hold cash only.
\end{itemize}

A graphical visualization of pair trading is presented in
Figure~\ref{fig:pairtrading}. Figure~\ref{fig:pairtrading}a shows the
market interactions according to the Spread (S) and thresholds. A position is opened whenever the spread deviates beyond the open threshold. The
position closure happens when the spread reverts below the close
threshold. Figure~\ref{fig:pairtrading}b, which shares the same time axis with (a), is a stretched view of (a). It presents the corresponding actions with the crossing of Spread (S) and zones. The spread deviations are classified into zones based on the Spread ($S$), Open-Threshold ($OT$),
and Close-Threshold ($CT$) Equation~(\ref{equ:zones}):

\vspace{-9pt}
\begin{figure}[H]
  %\centering
  \includegraphics[width=0.95\textwidth]{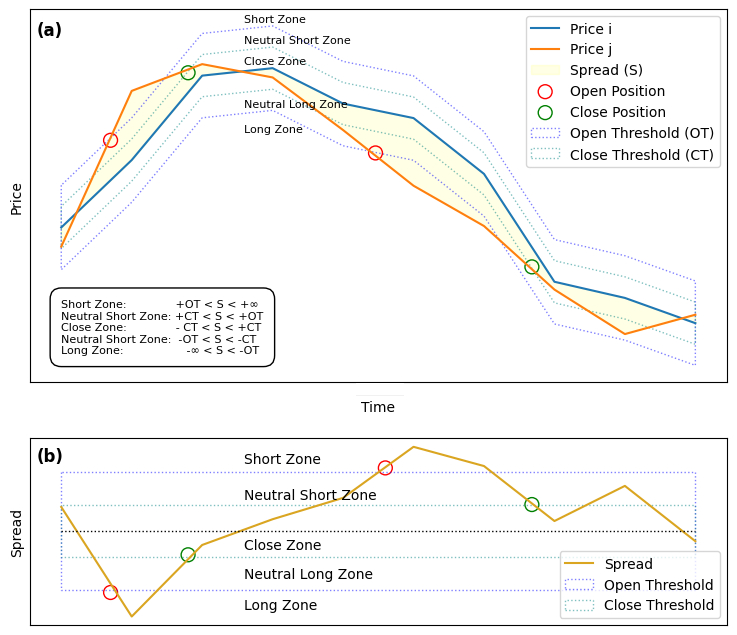}
  \caption{\label{fig:pairtrading}Stretched pair trading view of price distance 
 between $p_i$ and $p_j$. Figure (\textbf{b}), which shares the same time axis with (\textbf{a}), is a stretched view of (\textbf{a}). It presents the corresponding same actions with the crossing of Spread (S) and zones in two different views.}
\end{figure}

\begin{equation}
\left\{
\begin{array}{rl}
\text{Short Zone:} & +OT < S < +\infty, \\
                   & \text{spread deviates beyond open threshold} \\
\text{Neutral Short Zone:} & +CT < S < +OT, \\
                           & \text{spread deviates between open and close threshold} \\
\text{Close Zone:} & -CT < S < +CT, \\
                   & \text{spread reverts between close thresholds} \\
\text{Neutral Long Zone:} & -OT < S < -CT, \\
                          & \text{spread deviates between open and close threshold} \\
\text{Long Zone:} & -\infty < S < -OT, \\
                  & \text{spread deviates below open threshold} \\
\end{array}.
\label{equ:zones}
\right.
\end{equation}

\subsection{\acrlong{rl}}
\label{sec:acrlongrl}

\acrfull{rl} is used to train an agent to maximize rewards while
interacting with an environment~\citep{sutton_reinforcement_2018}. The
environment for \acrshort{rl} is required to be a
\acrfull{mdp}~\citep{bellman_markovian_1957}, which means it is modeled
as a decision-making process with the following elements〈State ($S$),
Action ($A$), Transition ($P_A$), Reward ($R_A$)⟩. The goal is to train
the agent to develop a policy ($\pi$) that fulfills an objective, e.g.,
maximizing profits in a trade. At every trading interval $t$, according
to state $S$, that the agent observes, action $A$ is chosen based
on policy $\pi$. The environment rewards/punishes the state transition
of $S_{t}\rightarrow S_{t+1}$ with environment reward $r$. If we assume
$\gamma$ to be the discount factor for the time-value discount of future
reward, \acrshort{rl} trains a policy $\pi$ that maximizes the total
discounted reward $G_t$, as shown in Equation~(\ref{equ:disctrwd}):

\begin{equation}
\label{equ:disctrwd}
G_t=\sum_{i=0}^{\infty}\gamma^i r_{t+i}.    
\end{equation}

\acrshort{rl} algorithms can broadly be classified based on three
criteria: value/policy-based, on/off policy, and actor/critic network
\citep{almahamid_reinforcement_2021}. Value-based methods estimate
state-action value functions for decision-making. Policy-based methods
directly learn action selection policies. The on-policy method requires
data generated by the current policy, and the off-policy method is capable of
leveraging past experiences from potentially different policies.
Moreover, actor--critic architectures, where an actor--network proposes
actions and a critic network evaluates, have shown a better performance
in facilitating policy improvement through this feedback loop. Most
recent research favors an actor--critic architecture instead of actor-only
or critic-only methods for better
performance~\citep{meng_reinforcement_2019, zhang_deep_2020}. Therefore,
only actor--critic algorithms are adopted in this~study. 

Based on the \acrshort{rl} classification criteria, some
representative algorithms have been selected for this study including
\acrfull{dqn} \citep{mnih_playing_2013}, \acrfull{sac}
\citep{haarnoja_soft_2019}, \acrfull{a2c}
\citep{sutton_reinforcement_2018}, \acrfull{ppo}
\citep{schulman_proximal_2017}. Diversified \acrshort{rl} algorithms
are experimented with to choose the most effective one in pair
trading. 

\section{Related Work\label{sec:relatedwork}}

\subsection{\acrlong{rl} in Algorithmic Trading}
\label{sec:acrl-algor-trad}

Reinforcement learning in AlphaGo captured the world's attention in
2016 by participating in a series of machine versus human competitions
on the board game GO \mbox{\citep{david_silver_demis_hassabis_alphago_2016}.}
Surprisingly, the research regarding RL in the financial market started
long before that. Recurrent reinforcement learning studies were the
mainstream works
\citep{gold_fx_2003,bertoluzzo_making_2007,maringer_regime-switching_2012,zhang_using_2016}
in the early stage of financial trading. After the upsurge of AlphaGo,
some significant advancements were brought to \acrshort{rl} trading as
well; \cite{huang_financial_2018} re-described
the \acrfull{mdp}
financial market as a game process to incorporate \acrshort{rl} as
a \textit{financial trading game
} \citep{huang_financial_2018}. 
\cite{pricope_deep_2021} proposed deep RL agents to develop profitable high-frequency trading strategies with sequential model-based optimization tuning the hyperparameters. With the recent development, newer \acrshort{rl} models such as \acrfull{dqn}, \acrfull{pg}, and \acrfull{a2c} have also been introduced by
researchers~\citep{meng_reinforcement_2019, zhang_deep_2020, mohammadshafie_deep_2024} for financial trading. A noteworthy research work is that of the FinRL group in the infrastructures and ensemble learning mechanism
%\citep{liu_finrl_2021, liu_finrl_2022, liu_finrl-meta_2022}. 
(\citeauthor{liu_finrl_2021} \citeyear{liu_finrl_2021}, \citeyear{liu_finrl_2022}, \citeyear{liu_finrl-meta_2022}).

\subsection{\acrlong{rl} in Pair Trading \label{sec:lrpairtrade}}

\acrlong{rl}, in combination with pair trading, is not an untapped
domain. \acrshort{rl} has ameliorated multifaceted aspects of the
traditional method of pair trading brought up by
\cite{gatev_pairs_2006}. The \acrshort{rl}
technique \acrlong{optics} contributed to the pair selection stage by
leveraging a clustering algorithm to produce better pair choices
\citep{sarmento_enhancing_2020}. \cite{vergara_deep_2024} traded deep reinforcement learning in the cryptocurrency market with ensemble practice, combining classical pair trading with the RL framework. \acrfull{credit},
an algorithm that takes into consideration both profitability and risks, was
engineered by \cite{han_mastering_2023}. Reward shaping is also an interesting
area where some work has been done in \acrshort{rl} trading
\citep{lucarelli_deep_2019, wang_improving_2021}.
\citeauthor{kim_optimizing_2019}'s work in~(\citeyear{kim_optimizing_2019})
is the most recent RL pair trading method. Their focus is on utilizing
RL to find the most trading opportunities. Instead of fixed thresholds, the
RL agent in Kim and Kim's work produces
thresholds for the upcoming trading period. Open, close, and stop-loss
thresholds determine the profits of pair trading.

Our work introduces a novel method to combine \acrshort{rl} with pair
trading. The work of \cite{gatev_pairs_2006} is not efficient
enough for a high-frequency market. The state-of-the-art method
of~\cite{kim_optimizing_2019} has some deficiencies: (i) it
requires the market's volatility to be relatively stable. The RL agent
may produce unsuitable thresholds if the market experiences increased
volatility. (ii) It lacks flexibility in the investment amount.
Opportunities with different qualities are programmed to be invested
with the same amount of capital. Once the RL agent determines a
threshold, the trading algorithm executes a trade at pre-determined thresholds. We leverage \acrshort{rl} to make investment
timing and quantity decisions. The adjustable investment amount is a
novel feature of our \acrshort{rl} pair trading. An \acrshort{rl} agent
measures how well the investment opportunities are based on observations
and invests a larger amount in more promising market conditions. Having
another dimension on the investment side should further enhance
profitability and reduce risks.

\section{Methodology \label{sec:methodology}}

\begin{figure}[H]
  %\centering
  \includegraphics[width=0.9\textwidth]{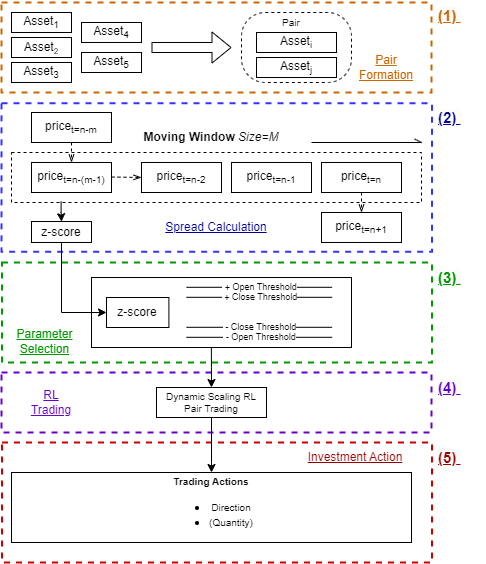}
  \caption{Architecture of trading strategies. }
  \label{fig:architecture.png}
\end{figure}

In this section, we introduce the architecture of the methodology
(Figure~\ref{fig:architecture.png}). The architecture includes five
steps: (1) \textit{pair formation
} for selecting assets to form a
tradeable pair (Section~\ref{sec:pairform}); (2) \textit{spread
  calculation 
} utilizing the moving-window technique to extract the
spread in a limited retrospective time frame (Section~\ref{sec:spread});
(3) \textit{parameter selection 
} from an historical dataset to decide the
most suitable hyperparameters for pair trading
(Section~\ref{sec:spread}); (4) \textit{\acrshort{rl} trading  
} by
allowing \acrshort{rl} to decide the trading timing and quantity in pair
trading (Section~\ref{sec:dcrl}); (5) \textit{investment action 
} for
taking the actions produced from \acrshort{rl} trading into market
execution.

\subsection{Pair Formation\label{sec:pairform}}

Pairs are selected based on two criteria: correlation and
cointegration. The widely adopted Pearson's correlation
\citep{perlin_m_2007, do_does_2010} is given by

\begin{equation}
\rho_{X,Y} = \frac{\mathrm{cov}(X,Y)}{\sigma_X \sigma_Y},
\end{equation}

\noindent
where \(\rho_{X,Y}\) is the correlation coefficient between assets \(X\)
and \(Y\), \(\mathrm{cov}(X,Y)\) is the covariance of \(X\) and \(Y\),
and \(\sigma_X\) and \(\sigma_Y\) are the standard deviations of \(X\) and
\(Y\), respectively. The Engle--Granger cointegration
test~\citep{burgess_using_2003, dunis_cointegration_2005} involves two
steps. First, the linear regression is performed:

\begin{equation}
  Y_t = \alpha + \beta X_t + \epsilon_t,
\end{equation}

\noindent
where \(Y_t\) and \(X_t\) are the asset price series, \(\alpha\) and
\(\beta\) are the regression coefficients, and \(\epsilon_t\) is the residual term.
The second step tests the residuals \(\epsilon_t\) for stationarity using an
\acrfull{adf} \citep{dickey_distribution_1979} test. The \acrshort{adf} test
regression is given in Equation~(\ref{eq:1}):
\begin{equation}
  \Delta \epsilon_t = \gamma \epsilon_{t-1} + \sum_{i=1}^{p} \delta_i \Delta \epsilon_{t-i} + \nu_t,
  \label{eq:1}
\end{equation}

\noindent
where \(\Delta \epsilon_t\) is the first difference of the residuals,
\(\gamma\) is the coefficient to be tested for stationarity, \(p\) is the
number of lagged difference terms included, and \(\nu_t\) is the error
term. If \(\gamma\) is significantly different from zero, the residuals are
stationary, indicating co-integration.

\begin{figure}[H]
  %\centering
  \includegraphics[width=0.8\textwidth]{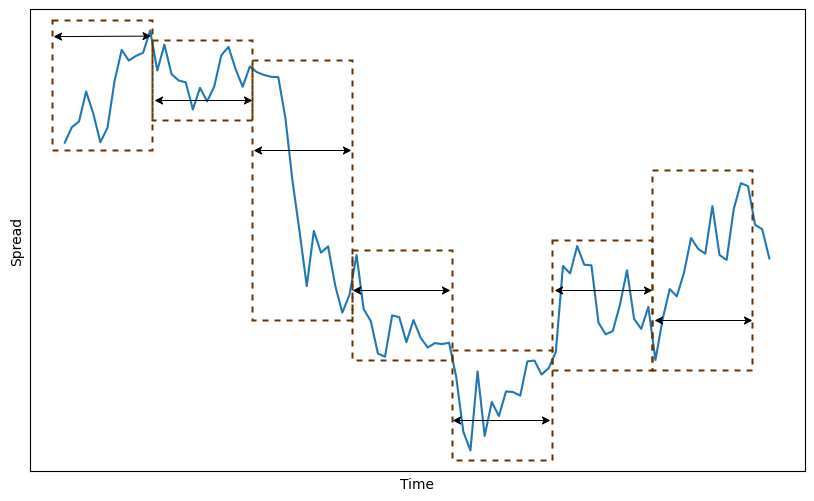}
  \caption{Window-size cut for correlation and co-integration testing.}
  \label{fig:pairformation.png}
\end{figure}

A moving window is applied to historical pricing data, as shown in Figure~\ref{fig:pairformation.png}. In this figure, the blue line represents the historical prices, while the dashed boxes illustrate the moving window. During the selection phase, averaged correlation and co-integration batches are employed to ensure that the selected assets exhibit a strong, long-term statistical~relationship.

\subsection{Spread Calculation \label{sec:spread}}

The second step is a moving window mechanism to capture the spread
movement (Figure~\ref{fig:architecture.png} Step-2). Spread
$\epsilon_{t}$ is calculated at every selected trading interval (e.g., every
five minutes) and is the error term $s$ from a regression between the two
prices $p_i$ and $p_j$. $\beta_0$ and $\beta_1$ are the intercept and slope, respectively, which follow a normal distribution with a mean of 0
and a standard deviation of $\sigma$ Equation~(\ref{equ:spreadcalc}):
\begin{equation}
  p_i = \beta_0 + \beta_1\times p_j + s_i\sim N(0, \sigma^2).
  \label{equ:spreadcalc}
\end{equation}

We normalize the spread with $z$-score Equation~(\ref{equ:zscore}) to
scale the spread into constant mean and standard deviation. The mean of
the spread in the sliding window is represented as $\bar{s}$:
\begin{equation}
  Z = \frac{{s - \bar{s}}}{{\sigma_s}} \label{equ:zscore}.
\end{equation}

% \subsection{Window-Sliding Pair Trading\label{sec:gridsearch}}

\subsection{Parameter Selection \label{sec:wspt_training}}

Three parameters to be explored are {window size}, {open threshold}, and {close threshold}. \textit{Window size $\in \mathbb{Z}^+$} is the number of historical samples in the moving window. \textit{Thresholds $ \in \mathbb{Q}^+$} are the entry and exit signals of trading actions that are highly linked to market \mbox{conditions}. 

Excessively wide thresholds suit more volatile markets, and conservatively narrow thresholds result in smaller but steadier wins. The combination of parameters of the
highest profitability 〈\textit{Window Size
}, \textit{Open Threshold
},
\textit{Close Threshold
}⟩ are selected from a search pool through a grid search in practice. \citeauthor{gatev_pairs_2006} \citeyearpar{gatev_pairs_2006} adopted 2 times the
standard deviation as the open threshold and the deviation
crossing point as the close threshold (Figure~\ref{fig:pairtrading}).
However, on the one hand, the parameters should vary with the arbitrage instruments as well as the market condition. Hence, the window-sliding mechanism is incorporated to reflect the heterogeneity of the pricing variance \citep{mandelbrot_variation_1967}.

\subsection{Reinforcement Learning Pair Trading \label{sec:dcrl}}

After we run the grid search of window-sliding pair trading, the next
problem concerns ``when'' and ``how much'' to trade. Pair trading
results from following pre-set rules (à la \citeauthor{gatev_pairs_2006}~\citeyear{gatev_pairs_2006}), which are obtainable using window-sliding
pair trading. However, we want to know if \acrshort{rl} produces better
investment decisions than blindly following the rules. Therefore, the
most profitable parameter combination is passed onto further
\acrshort{rl}-based pair trading so that we can compare the results
between \acrshort{rl}-based pair trading and non-\acrshort{rl} pair~trading. 

\subsubsection{Observation Space}
Observation space stands for the information an \acrshort{rl} agent
observes. The agent observes market information to make decisions.
The observations adopted for our \acrshort{rl} environment are as
follows: 〈\textit{Position}, \textit{Spread}, \textit{Zone}⟩.

\begin{figure}[H]
    % \centering
     \includegraphics[width=0.8\textwidth]{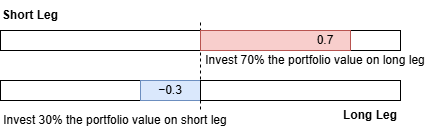}
     \caption{The value
 of position observation based on investment.}
     \label{fig:actionspace.png}
   \end{figure}

 \begin{itemize}
 \item \textbf{Position $\in [-1, 1]$}: Position stands for the current
   portfolio value. Position is a percentage measuring the direction of
   investment (c.f. Figure~\ref{fig:actionspace.png}). As in pair trading, we define longing the first asset with shorting the second asset as `holding a long leg' and the other way around as `holding a short leg'. Assuming that we
   do not use leverage, holding a long leg with a 70\% portfolio value
   gives Position = 0.7. Holding a short leg with a 30\% portfolio value
   gives Position = $-$0.3. Position 0 means we only hold cash.

 \item \textbf{Spread $\in \mathbb{R}$}: This represents how much the current spread
   has deviated from the mean (Section~\ref{sec:spread}).
 
 \item \textbf{Zone $\in \{\text{Zones}\}$}: Zone is an important
   indicator that comes from the comparison between the $z$-score with
   the thresholds for signals (Figure~\ref{fig:pairtrading}b).
   Traditional pair trading \citep{gatev_pairs_2006, yang_optimal_2024}
   takes the zone as the direct trading signal. However, in \acrshort{rl}-based
   pair trading, the zone is an observation for the \acrshort{rl} agent
   to make better decisions.
 \end{itemize}
\vspace{-11pt}
\subsubsection{Action Space}
\label{sec:action-space}

Since the pair trading technique is a relatively low-risk strategy, most
applications invest with a fixed amount or the complete portfolio
value~\citep{burgess_using_2003, perlin_evaluation_2009,
  huck_pairs_2010}. Meanwhile, it is natural for an experienced trader
to invest different amounts based on the quality of opportunities.
Opportunities with a higher probability of success are worth more
investment. Therefore, we investigate granting the \acrshort{rl} agent
not only the decision of when to invest but also the freedom to choose
the investment amount. 

Take $A \in [-1, 1]$ as the action. Similar to the observation space (c.f.
Figure~\ref{fig:actionspace.png}), the action ranges from $-$1 to 1,
representing the investment as a percentage of the portfolio value in the
long leg and short leg directions. Investing 50\% of the portfolio as a
long leg means A = 0.5. Investing 20\% of the portfolio in the short leg
means A= $-$0.2. 

In practice, we have to consider the relationship between the existing
position and the next action. We classify the execution of action
as {\textit{open position}
}, {\textit{adjust position}
}, or \textit{close~position}
.
\begin{itemize}
\item \textbf{Open position} is the action of opening a new position.
\item \textbf{Close position} is the closure of a position.
\item \textbf{Adjust position} happens when a previous position is open,
  and the \acrshort{rl} agent wants to open another position. For
  example, if the current position is a 70\% long-leg and the new action
  is A = 0.8, \textit{only} the extra 10\% shall be actioned.
\end{itemize}

\subsubsection{Reward Shaping}
\label{sec:reward-shaping}

The \acrshort{rl} reward consists of three components: {\textit{action
  reward}
}, {\textit{portfolio reward}
}, and {\textit{transaction punishment}
}.

\begin{itemize}
\item \textbf{Portfolio reward
} is the profit/loss from closing a
  position. The portfolio value $V_p$ only updates when the position closes.
  $V'_{p}$ is the position value at the start of a trading period
  $p$. Then, upon closing the trade at the end of trading period $p$,
  the reward it calculated, as shown in
  Equation~(\ref{equ:portfoliorwd}).
  \begin{equation}
    \label{equ:portfoliorwd}
    \text{Profit Reward} = V_p - V'_{p}.
  \end{equation}

\item \textbf{Action reward 
} means the agent needs to be rewarded for
  taking a desired action in the corresponding zone. In general, the
  agent is free to decide on any action. However, we use \textit{action
    reward 
} to encourage the agent to choose desired actions. It rewards the agent for making a desired
  action in certain zones (Table~\ref{tab:zone-action}) with some
  freedom in neutral zones. The stronger the action reward, the more it
  resembles traditional pair~trading.

\item \textbf{Transaction punishment} is a negative reward for
  encouraging small adjustments instead of large changes in the
  position. The punishment is the difference between the action and
  position. If the current position in observation is $P$ and the action
  is $A$, the transaction punishment is Equation~(\ref{equ:txpunish}):
  \begin{equation}
    \label{equ:txpunish}
    \text{Transaction Punishment} = P-A.
  \end{equation}
\end{itemize}
\unskip
  \begin{table}[H]
    
    \caption{Rewarding
 behaviors in zones.\label{tab:zone-action}}
    \small\setlength{\tabcolsep}{20.5mm}\begin{tabular}{ll}
      \toprule
      \textbf{Zones}              & \textbf{Rewarding Behavior} \\ \midrule
      Short Zone         & Short leg               \\
      Neutral Short Zone & Short leg or Close        \\
      Close Zone         & Close               \\
      Neutral Long Zone  & Long leg or Close         \\
      Long Zone          & Long leg                \\ \bottomrule
    \end{tabular}
  \end{table}
  \unskip

\subsection{Dynamic Agents}

Trading agents in RL environments operate by taking actions within a defined action space based on observations from the state space. This section details the design of two dynamic agent settings, RL$_1$ and RL$_2$, each tailored to address specific aspects of pair trading. These agents leverage the Markov Decision Process (MDP) framework to model the complexities of financial markets, enabling adaptive decision-making in volatile conditions.

In the first setting, RL$_1$ is tasked with optimizing trade timing and directionality. Pair trading is modeled as a Markov Decision Process (MDP), represented as \((\mathcal{S}_1, \mathcal{A}_1, \mathcal{T}_1, r_1, \gamma_1)\):
\begin{itemize}
    \item \(\mathcal{S}_1\) represents the state space, including normalized price spreads, historical z-scores, and zone indicator.
    \item \(\mathcal{A}_1\) defines a discrete action space consisting of three possible actions: initiating a long--short position, closing existing positions, or initiating a short--long position. This allows the agent to determine the optimal direction and timing for trades.
    \item \(r_1(s, a)\) is the reward function, defined as follows:
    \[
    r_1(s_t, a_t) = \begin{cases} 
    \Delta P_t - c, & \text{if a trade is executed;} \\ 
    0, & \text{if no trade is executed,}
    \end{cases}
    \]
    where \(\Delta P_t\) represents the profit or loss from the trade, and \(c\) is the transaction cost. The reward function penalizes the agent for transaction costs while directly linking rewards to trade profitability.
    \item The goal is to maximize the cumulative discounted reward:
    \[
    R_1 = \mathbb{E} \left[\sum_{t=0}^\infty \gamma_1^t r_1(s_t, a_t)\right].
    \]
\end{itemize}

RL$_2$ extends RL$_1$ by shifting focus from trade timing to determining the investment quantity for a given trade opportunity. It models pair trading as an MDP defined by \((\mathcal{S}_2, \mathcal{A}_2, \mathcal{T}_2, r_2, \gamma_2)\):
\begin{itemize}
    \item \(\mathcal{S}_2\) is the state space, which is identical to RL$_1$.
    \item \(\mathcal{A}_2 = [-1, 1]\), where the continuous value represents the investment quantity. Here, \(0\) stands for no involvement, positive values represent buying, and negative values represent selling.
    \item \(r_2(s, a)\) is the reward function, defined as follows:
    \[
    r_2(s_t, a_t) = \Delta P_t \cdot a_t - c(|a_t|),
    \]
    where \(c(|a_t|)\) represents transaction costs proportional to the absolute investment size \(|a_t|\). This reward structure incentivizes the agent to optimize both the direction and magnitude of its investment.
    \item The objective is to maximize the cumulative discounted reward:
    \[
    R_2 = \mathbb{E} \left[\sum_{t=0}^\infty \gamma_2^t r_2(s_t, a_t)\right].
    \]
\end{itemize}

The primary differences between RL$_1$ and RL$_2$ lie in their action spaces and reward functions. RL$_1$ operates with a discrete action space and focuses on optimizing directional timing and trade management. In contrast, RL$_2$ uses a continuous action space \(\mathcal{A}_2 = [-1, 1]\), enabling it to adjust investment sizes dynamically. The environments are designed to guide the agents by rewarding profitable actions and penalizing costly ones, encouraging effective decision-making for timing and quantity. The exact mechanisms driving these decisions are embedded within the neural network, shaped by the agent's interactions with the~environment.

\section{Benchmark Results \label{sec:experiments}}

Next, we carry out experiments using the proposed methodology. We adopt
the same dataset and the same parameters for non-\acrshort{rl} pair
trading and \acrshort{rl} pair trading for comparison purposes.

\subsection{Experimental Setup}
\label{sec:experimental-setup}

We experiment with window-sliding pair trading and
\acrshort{rl} pair trading in the cryptocurrency market. The
cryptocurrency market is famous for its volatility, easy access, and 24/7
operating time. Data preprocessing follows two steps. Firstly, the evaluation of the correlations among trading pairs for selection of arbitrage candidates; next, searching through the combination of thresholds and retrospective periods for suitable signals.

\subsubsection{Datasets}
\label{sec:datasets}

\begin{figure}[H]
  %\centering
  \includegraphics[width=0.8\textwidth]{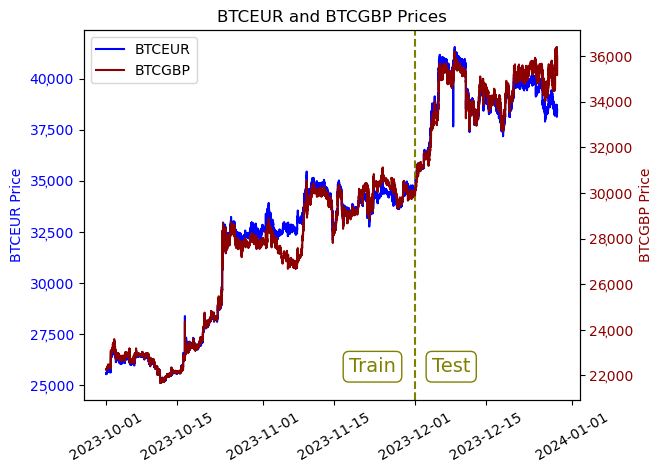}
  \caption{Prices
 of BTCEUR and BTCGBP.}
  \label{fig:traintest}
\end{figure}

The application of our trading methodology is on Binance, the largest
cryptocurrency market.\endnote{\url{https://www.binance.com/en}, accessed on 8 November 2024.
} For the best
market liquidity, we picked Bitcoin--Fiat currencies under different
trading intervals for pair trading. Pair formation criteria are based on
Pearson's correlation and augment the Engle--Granger two-step cointegration
test (Section~\ref{sec:pairform}) for quote currencies that follow a similar trend against the base currency (Figure~\ref{fig:traintest}). 
The formation period is from October 2023 to November 2023, and the test is in December 2023, with trading intervals of 1 min
(121,500 entries), 3 min (40,500 entries), and 5 min (24,300 entries),
respectively. We exhaustively compared correlation and co-integration for
the best pair (Table~\ref{tab:corr_coint}).\endnote{While calculating
  the co-integration and correlation, intervals with low volume trades
  are exempted from the calculation.} Although Binance has quite a few
fiat currencies, only the US Dollar (USD), Great British Pound (GBP),
Euro (EUR), and Russian Ruble (RUB) display relatively strong liquidity.
The pair with the strongest correlation and co-integration is BTCEUR and
BTCGBP under a 1~min trading interval~(Table \ref{tab:corr_coint}). 

\unskip
\begin{table}[H]  
  \caption{Correlation
 and co-integration of pair formation.}
 \small\setlength{\tabcolsep}{4.3mm} \begin{tabular}{lcccccc}
    \toprule
    \textbf{Pairs} & \multicolumn{2}{l}{\textbf{1m}} & \multicolumn{2}{l}{\textbf{3m}} & \multicolumn{2}{l}{\textbf{5m}} \\
                   & \multicolumn{1}{l}{\textbf{coint}} & \multicolumn{1}{l}{\textbf{corr}} & \multicolumn{1}{l}{\textbf{coint}} & \multicolumn{1}{l}{\textbf{corr}} & \multicolumn{1}{l}{\textbf{coint}} & \multicolumn{1}{l}{\textbf{corr}} \\
    \midrule
    BTCEUR-BTCGBP & 0.5667         & 0.8758         & 0.4667         & 0.8759         & 0.4667         & 0.8754         \\
    BTCEUR-BTCRUB & 0.3333         & 0.8417         & 0.3333         & 0.8417         & 0.3167         & 0.8416         \\
    BTCEUR-BTCUSD & 0.1667         & 0.9328         & 0.2000         & 0.9327         & 0.2000         & 0.9329         \\
    BTCGBP-BTCRUB & 0.3500         & 0.7606         & 0.3333         & 0.7608         & 0.3333         & 0.7603         \\
    BTCGBP-BTCUSD & 0.4833         & 0.8404         & 0.4167         & 0.8403         & 0.4000         & 0.8403         \\
    BTCRUB-BTCUSD & 0.4000         & 0.8538         & 0.3333         & 0.8539         & 0.3500         & 0.8543 \\
    \bottomrule
  \end{tabular}
  \label{tab:corr_coint}
\end{table}

The transaction cost in the experiment is set to 0.02\% commission based on
Binance's fee
scheme.\endnote{\url{https://www.binance.com/en/fee/futureFee}, accessed on 8 November 2024.}
 The transaction cost of 0.02\% is a flat percentage charge for transactions
in both directions. A pair trading leg, including long, the first asset,
and short, the second asset, is charged for both long and short actions.

\subsubsection{Grid Search and Reinforcement Learning}
\label{sec:grid-search-reinf}

A grid search is used to find a set of profitable parameters, including the
\textit{open/close threshold
} and \textit{window size
}during the
training period (October 2023 to November 2023). Every iteration of the
window-sliding pair trading experiments has one set of parameters
(window size, open/close threshold) until exhaustion. We start off the exploration based on the experiential estimation of the asset characteristics. The most
profitable parameter set will be used to test traditional pair trading
during the test period (December 2023) and also for testing the proposed
\acrshort{rl} strategy.

The profitability in conducting a grid search is measured by the Total Compound Return
(RTOT), where $V_p$ and $V_p'$ are the values of the portfolio at the
beginning of the period and the end of the period, and $t$ is the total
length of the trading period in Equation~(\ref{equ:rtot}):
\begin{equation}
    \label{equ:rtot}
    rtot = (V_p' / V_p)^{1/t}-1 \times 100\%.
\end{equation}

During the training period, the most profitable parameter set
is \textit{open threshold
} = 1.8 $z$-score, \textit{close threshold
} =
0.4 $z$-score, and \textit{window size
} = 900 intervals. Some example
results of the grid search are presented in
Table~\ref{tab:gridsearch}.

\begin{table}[H]
\caption{Trading parameter tuning.}
\label{tab:gridsearch}
\small\setlength{\tabcolsep}{8.45mm}\begin{tabular}{llll}
 \toprule
\textbf{OPEN
\_THRES} & \textbf{CLOS\_THRES} & \textbf{PERIOD} & \textbf{RTOT (\%)} \\ 
\midrule
4.0 & 2.0 & 2000 & 0.0651 \\
4.0 & 0.5 & 500  & 0.5024 \\
3.0 & 1.0 & 500  & 0.9993 \\
3.0 & 0.5 & 1000 & 0.8932 \\
3.0 & 0.5 & 500  & 1.0704 \\
2.5 & 0.3 & 700  & 2.1542 \\
2.5 & 0.5 & 700  & 1.5667 \\
3.0 & 0.3 & 700  & 1.3160 \\
2.1 & 0.4 & 700  & 2.5633 \\
2.1 & 0.3 & 800  & 2.6916 \\
2.3 & 0.4 & 800  & 2.3096 \\
2.1 & 0.4 & 800  & 2.8202 \\
2.0 & 0.4 & 1000 & 2.7339 \\
2.0 & 0.4 & 900  & 3.0400 \\
1.9 & 0.3 & 900  & 2.8989 \\
1.9 & 0.4 & 900  & 3.1077 \\
1.8 & 0.4 & 900  & 3.0565 \\
... & ... & ...  & ...    \\
\bottomrule
\end{tabular}
\end{table}

The setup of \acrshort{rl}-based pair trading relies on these
parameters. The window size decides the retrospective length of the
spread, and the thresholds decide the zones. Algorithms such as \acrshort{ppo} and \acrshort{a2c} are applicable to both discrete and continuous action spaces. Some algorithms, e.g.,
\acrshort{dqn}, can only be used in discrete space, and \acrshort{ddpg}
is only applicable in a continuous space. Therefore, we adopt
\acrshort{ppo}, \acrshort{dqn}, and \acrshort{a2c} in \acrshort{rl} pair
trading, which decide the timing, and \acrshort{ppo},
\acrshort{a2c}, and \acrshort{sac} in \acrshort{rl} pair trading, which decide both the timing and investment quantity. The algorithms are adopted from the Baseline3 collection \citep{raffin_stable-baselines3_2021}.

\subsubsection{Evaluation Metrics}

Our main concern is the highest profitability in trading techniques. We care about the cumulative return, which is the total profit for the testing period, as well as the annualized return \acrfull{cagr}. With $V(t_0)$ as the initial state, $V(t_n)$ as the final state, and $t_n - t_0$ as the period of trading in years, the \acrshort{cagr} is Equation~(\ref{equ:cagr}):

\begin{equation}\label{equ:cagr}
\operatorname{CAGR}\left(t_0, t_n\right)=\left(\frac{V\left(t_n\right)}{V\left(t_0\right)}\right)^{\frac{1}{t_n-t_0}}-1.
\end{equation}

There are some popular indicators for distinguishing whether a strategy
is profit--risk effective, e.g., the Sharpe ratio. In the Sharpe ratio
\citep{sharpe_capital_1964}, $R_p$ is the return of the trading
strategy, $R_f$ is the interest rate,\endnote{We adopt the Federal
  Reserve interest rate of 5.5\%, which is correct as of 13 June 2024.} and $\sigma_p$ is the
standard deviation of the portfolio's excess return
(Equation~(\ref{equ:sharperatio})):
\begin{equation}\label{equ:sharperatio}
    \text{Sharpe Ratio}=\frac{R_p - R_f}{\sigma_p}.
\end{equation}

We also care about the strategies' activities, such as the order count and win/loss ratio. The indicators used for comparison are presented in Table~\ref{tab:metricsdesc}.

\begin{table}[H]
  \caption{Descriptive table of evaluation metrics.}
  \label{tab:metricsdesc}  
  \small\setlength{\tabcolsep}{11.2mm}\begin{tabular}{ll}
    \toprule
    \textbf{Profitability Indicator} & \textbf{Description} \\
    \midrule
    Cumulative Return & Profit achieved during trading period \\
    CAGR & Compound Annual Growth Rate \\
    Sharpe Ratio & Risk-adjusted returns ratio \\
    \midrule
    \textbf{Activity Indicator} & \textbf{Description} \\
    \midrule
    Total Action Count & Total orders executed \\
    Win/Loss Action Count & Number of winning/losing trades \\
    Win/Loss Action Ratio & Ratio of winning to losing trades \\
    Max Win/Loss Action & Maximum profit/loss per Action in Bitcoin \\
    Avg Action Profit/Loss & Average profit/loss per trade in Bitcoin \\
    Time in Market & Percentage of time invested in the market \\
    \midrule
    \textbf{Risk Indicator} & \textbf{Description} \\
    \midrule
    Volatility (ann.) & Annualized standard deviation of returns \\
    Skew & Asymmetry of returns distribution \\
    Kurtosis & ``Tailedness'' of returns distribution \\
    \bottomrule
  \end{tabular}
\end{table}

\subsection{Experimental Results}
\label{sec:experimental-results}

We present the profitability and risk results from our experiments along
with the trading indicators.

\subsubsection{Result Comparison}
\label{sec:result-comparison}

Our work is compared with standard pair trading
(Section~\ref{sec:tradpairtrade}) and state-of-the-art pair trading
techniques (Section~\ref{sec:lrpairtrade}).

Our results are presented in Table~\ref{tab:backtest}. The results
display a positive return for the traditional pair trading
technique~\cite{gatev_pairs_2006}. The algorithm \acrshort{a2c}
displays a positive return for \acrshort{rl} pair trading techniques.
However, the \acrshort{ppo}, \acrshort{sac}, and \acrshort{dqn} algorithms
do not perform as well as \acrshort{a2c}. If we view \acrshort{a2c} as
the chosen algorithm for pair trading, the results show a steady income
from pair trading. The traditional pair trading of \cite{gatev_pairs_2006} displays a stable income compared to
others due to its rule-based execution stability. 

The first adoption of \acrshort{rl}$_1$ pair trading is close to the traditional method. The results table shows that it achieved much better results than the traditional pair trading approach~\mbox{\cite{gatev_pairs_2006}}. The second adoption of
\acrshort{rl}$_2$ pair trading is significantly different from
\acrshort{rl}$_1$ trading, which decides only timing and produces more
profit than other techniques under the same level of volatility.
\citeauthor{kim_optimizing_2019}'s (\citeyear{kim_optimizing_2019})
method did not achieve a positive return. Since the method was developed
for the forex market, it has not adapted well to the extremely volatile
cryptocurrency world.
\startlandscape

\begin{table}[H]
 % \begin{threeparttable}
    \caption{Evaluation metrics comparison between trading techniques.    \label{tab:backtest}}
    \small\setlength{\tabcolsep}{3.84mm}\begin{tabular}{lllllllllll}
      \toprule
      & \textbf{\citeauthor{gatev_pairs_2006} (\citeyear{gatev_pairs_2006})} & \multicolumn{3}{c}{\textbf{\citeauthor{kim_optimizing_2019} (\citeyear{kim_optimizing_2019})}} & \multicolumn{3}{c}{\textbf{RL\boldmath{$_1$}}} & \multicolumn{3}{c}{\textbf{RL\boldmath{$_2$}}} \\
      \textbf{\acrshort{rl} Algo.} & \textbf{NA} & \textbf{\acrshort{ppo}} & \textbf{\acrshort{a2c}} & \textbf{\acrshort{dqn}} & \textbf{\acrshort{ppo}} & \textbf{\acrshort{a2c}} & \textbf{\acrshort{dqn}} & \textbf{\acrshort{ppo}} & \textbf{\acrshort{a2c}} & \textbf{\acrshort{sac}} \\
      \midrule
      \textbf{Profitability} &&&&&&&&&& \\
      Cumulative Return & 8.33\% & $-$0.16\% & $-$35.16\% & $-$35.79\% & 1.89\% & 9.94\% & $-$31.99\% & $-$77.81\% & 31.53\% & $-$87.12\% \\
      CAGR & 195.12\% & $-$2.19\% & $-$99.71\% & $-$99.75\% & 30.05\% & 278.72\% & $-$99.56\% & $-$100.00\% & 3974.65\% & $-$100.00\% \\
      Sharpe Ratio & 25.91 & $-$1.67 & $-$2.04 & $-$2.60 & 5.44 & 32.74 & $-$8.77 & $-$1.99 & 94.34 & $-$1.93 \\
      \midrule
      \textbf{Activities} &&&&&&&&&& \\
      Total Action Count & 490 & 43 & 1248 & 1062 & 1304 & 249 & 879 & 3443 & 229 & 2798 \\
      Won Action Count & 284 & 24 & 600 & 503 & 578 & 240 & 232 & 842 & 162 & 917 \\
      Lost Action Count & 206 & 19 & 648 & 559 & 726 & 9 & 647 & 2601 & 67 & 1881 \\
      Win/Loss Action Ratio & 1.38 & 1.26 & 0.93 & 0.90 & 0.80 & 26.67 & 0.36 & 0.32 & 2.42 & 0.49 \\
      Max Win Action (USD) & 75.35 & 163.52 & 606.75 & 606.75 & 43.72 & 121.74 & 70.59 & 307.78 & 648.87 & 160.15 \\
      Max Loss Action (USD) & $-$27.73 & $-$187.86 & $-$763.70 & $-$553.25 & $-$108.51 & $-$21.33 & $-$282.84 & $-$389.22 & $-$64.97 & $-$1456.43 \\
      Avg Win Action Profit/Loss (USD) & 14.50 & 41.72 & 38.68 & 37.47 & 5.51 & 17.77 & 11.30 & 15.88 & 90.94 & 8.03 \\
      Avg Loss Action Profit/Loss (USD) & $-$2.90 & $-$54.68 & $-$58.75 & $-$60.78 & $-$3.28 & $-$7.06 & $-$24.95 & $-$17.78 & $-$21.00 & $-$23.49 \\
      \midrule
      \textbf{Risk} &&&&&&&&&& \\
      Volatility (ann.) & 6.01\% & 3.93\% & 51.43\% & 40.43\% & 3.61\% & 6.30\% & 11.92\% & 53.04\% & 27.30\% & 54.66\% \\
      Skew & 1840 & $-$54 & $-$358 & $-$358 & $-$874 & 2673 & $-$1899 & $-$374 & 4314 & $-$3048 \\
      Kurtosis & 48,145 & 135 & 4138 & 4201 & 133,603 & 114,944 & 54,808 & 12,987 & 254,283 & 138,851 \\
      \bottomrule
    \end{tabular}
    
\begin{adjustwidth}{+\extralength}{0cm}
%\centering %% If there is a figure in wide page, please release command \centering
\noindent{\footnotesize{The trading period is from 1 December 2023 to 31 December 2023. The transaction cost is 0.02\%, and the interest rate is 5.5\%. \acrshort{rl}$_1$ stands for the pair trading that allows \acrlong{rl} to decide upon the investment timing. \acrshort{rl}$_2$ stands for \acrlong{rl} pair trading that allows the \acrshort{rl} agent to decide both investment timing and quantity.}} 
\end{adjustwidth}
  %\end{threeparttable}
\end{table}
\finishlandscape

Behavior-wise, \acrshort{ppo}, \acrshort{dqn}, and \acrshort{sac} tend to conduct excessive transactions that are not profitable. On the contrary, \acrshort{a2c} have fewer trades but higher profits on each trade. \acrshort{rl}$_2$ pair trading shows further fewer total actions because of the adjusted position action, where we do not consider a position adjustment as one trade until it is closed. Apart from the result in Table~\ref{tab:backtest}, the portfolio growth trend with the best-performing \acrshort{rl} algorithm agent is presented in Figure~\ref{fig:results} (a comparison with \cite{gatev_pairs_2006} is provided in Figure~\ref{fig:comparison}a in the Appendix). Most of the pair trading experiments, including \cite{gatev_pairs_2006}, RL$_1$, and RL$_2$, display a stable upturn, which is ideal from the perspective of pair trading. From the drawdown graphs, we can observe that \acrshort{rl}$_1$ produces fewer drawdowns compared to the non-RL pair trading method from \cite{gatev_pairs_2006} and has a significantly higher win/loss action ratio due to differences in threshold settings. However, RL$_1$’s cumulative profit is not consistently higher, and when transaction fees are zero, its cumulative profit is slightly lower than that of the Gatev et al. method. RL$_2$ displays the strongest profitability, despite a lower win/loss action ratio, due to its progressive trading strategy. In general, all three pair trading methods show the ability to generate stable income in a volatile trading~market.

\begin{figure}[H]
  \centering
  \begin{subfigure}[b]{0.497\textwidth}
    \includegraphics[width=\textwidth]{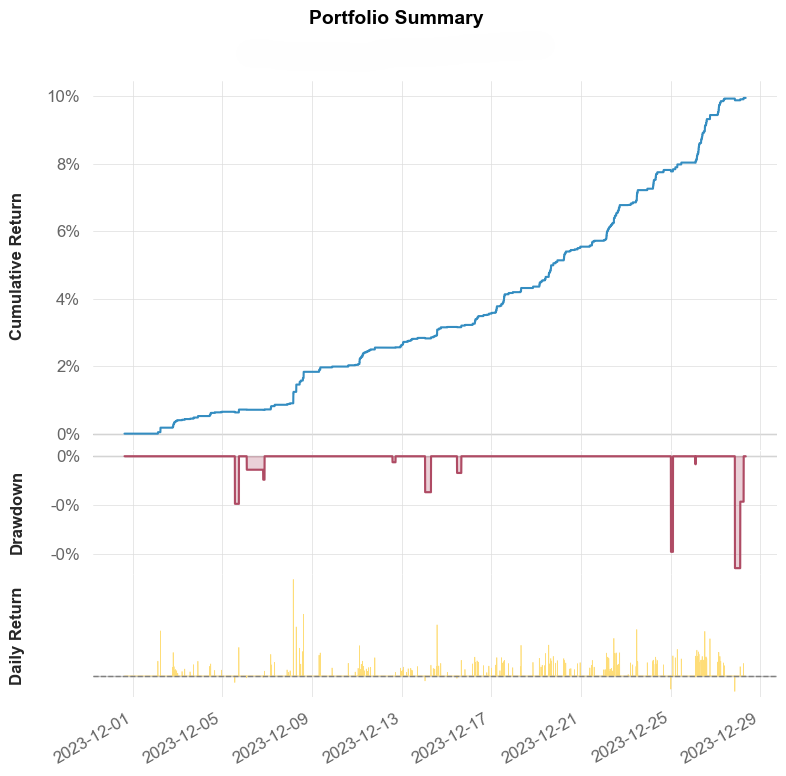}
    \caption{\centering\acrshort{rl}$_1$ Pair Trading (A2C).}
    \label{fig:fixedamtpt}
  \end{subfigure}
  \hfill
  \begin{subfigure}[b]{0.497\textwidth}
    \includegraphics[width=\textwidth]{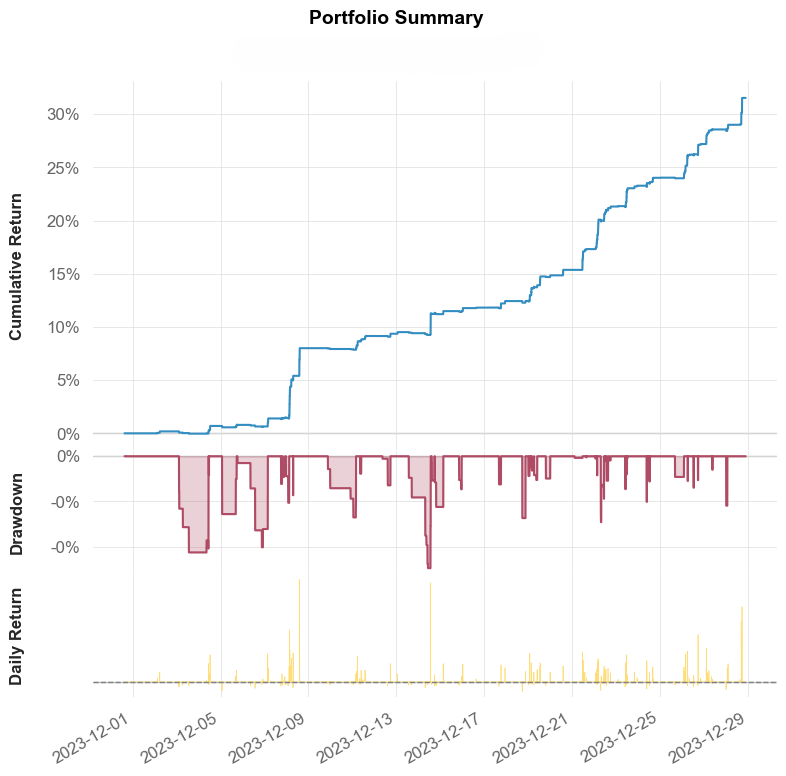}
    \caption{\centering\acrshort{rl}$_2$ Pair Trading (A2C).}
    \label{fig:freeamtpt}
  \end{subfigure}\vspace{+6pt}
  \caption{Comparison
  of portfolio value trends for \acrshort{rl}$_1$ and \acrshort{rl}$_2$ Pair Trading (A2C).}
  \label{fig:results}
\end{figure}

\subsubsection{Effect of Transaction Cost}
\label{sec:effect-trans-cost}

The profitability of high-frequency trading techniques is always
significantly impacted by transaction costs. Cryptocurrency exchanges
normally provide a large-volume discount scheme. The Binance fee ranges
from 0.02\% to even 0\%, depending on the volume and the holding of
their membership token. Considering that the users of these techniques
may benefit from different transaction fee tiers, we explore the trading
techniques under different transaction cost tiers as well.
With more exploration under 0.05\%, 0.01\%, and 0\% transaction costs
compared to the default 0.02\% transaction cost, we can see the
significant impact of decreasing the transaction cost. The participating
approaches are adopted with the most profitable algorithm based on the
backtesting result (Table~\ref{tab:backtest}). We can see that trading
techniques generally perform better under lower transaction costs.
\acrshort{rl}-based techniques tend to perform more trades when the
transaction costs are lower.

\section{Discussion and Conclusions\label{sec:conclusion}}

Pair trading has been a popular algorithmic trading method for decades.
The in-demand high-frequency trading domain requires a fast-track
decision-making process. However, the traditional rule-based pair
trading technique lacks the flexibility to cater to volatile market
movement. In this research, we proposed a mechanism to adopt
\acrfull{rl} to observe the market and produce profitable pair trading
decisions. The first adoption of \acrlong{rl} pair trading grants the \acrshort{rl}$_1$
agent the flexibility to decide action timing. The second adoption of \acrlong{rl}$_2$ pair trading further gives the \acrshort{rl} agent the
access to decide the timing and invest quantity. 

% Compare DSRL with Gatev and Kim
We compared it to the traditional rule-based pair trading technique
\citep{gatev_pairs_2006} and a
state-of-the-art \acrshort{rl} pair trading technique
\citep{kim_optimizing_2019} for
December 2023 in the cryptocurrency market for BTCEUR and BTCGBP under a standard
future 0.02\% transaction cost. Kim and Kim's method does not perform well in the cryptocurrency world. Gatev et al.'s method achieved 8.33\% per trading
period. Our first adoption of the \acrshort{rl}$_1$ method achieved 9.94\%, and the second adoption of the \acrshort{rl}$_2$ method achieved 31.53\% returns during the trading period. The outperformance is generally consistent across different
transaction costs. The evaluation metrics show that \acrshort{rl}-based
techniques are generally more active than traditional techniques in the
cryptocurrency market under various transaction costs. In general, our
trading methods have greater market participation
than~Gatev et al.'s traditional rule-based pair trading
and~Kim and Kim's threshold-adaptive \acrshort{rl}
pair trading (Tables~\ref{tab:backtest} and \ref{tab:robustness}).

\begin{table}[H]

\caption{Evaluation metrics comparison under different transaction costs.}
\label{tab:robustness}
%\begin{threeparttable}
\hspace{-9pt}
\scalebox{0.9}{
\small\setlength{\tabcolsep}{2.99mm}\begin{tabular}{lllll}
\toprule
\textbf{Indicators} & \multicolumn{4}{c}{\textbf{Trading Approaches}}\\
\midrule
 0.05\% Transaction Fee & \citeauthor{gatev_pairs_2006} (\citeyear{gatev_pairs_2006}) & \citeauthor{kim_optimizing_2019} (\citeyear{kim_optimizing_2019}) & RL{$_1$} & RL{$_2$}\\
\midrule
Cumulative Profit & 5.02\% & -0.26\% & 5.76\% & 7.40\% \\
Sharpe Ratio & 14.60 & -2.34 & 21.00 & 7.82 \\
Total Action Count & 490 & 43 & 154 & 207\\
Won Action Count & 246 & 23 & 152 & 110 \\
Lost Action Count & 244 & 20 & 2 & 97 \\
Win/Loss Action Ratio & 1.01 & 1.15 & 76.00 & 1.13\\
Max Win Action (USD) & 72.82 & 114.76 & 43.36 & 606.22\\
Max Loss Action (USD) & $-$30.26 & $-$169.52 & $-$8.30 & $-$168.81\\
Avg Win Action Profit/Loss (USD) & 13.57 & 37.94 & 16.10 & 70.99\\
Avg Loss Action Profit/Loss (USD) & $-$4.99 & $-$47.68 & $-$5.64 & $-$48.28\\
\midrule
 0.01\% Transaction Fee & \citeauthor{gatev_pairs_2006} (\citeyear{gatev_pairs_2006}) & \citeauthor{kim_optimizing_2019} (\citeyear{kim_optimizing_2019}) & RL{$_1$} & RL{$_2$}\\
\midrule
Cumulative Profit & 9.43\% & $-$1.13\% & 9.88\% & 33.99\% \\
Sharpe Ratio & 29.84 & $-$7.07 & 33.24 & 104.40 \\
Total Action Count & 490 & 43 & 251 & 181\\
Won Action Count & 317 & 20 & 242 & 149 \\
Lost Action Count & 173 & 23 & 9 & 32 \\
Win/Loss Action Ratio & 1.83 & 0.87 & 26.89 & 4.66\\
Max Win Action (USD) & 76.20 & 65.99 & 121.74 & 675.23\\
Max Loss Action (USD) & $-$26.88 & $-$169.66 & $-$21.33 & $-$27.91\\
Avg Win Action Profit/Loss (USD) & 13.93 & 24.88 & 17.52 & 98.74\\
Avg Loss Action Profit/Loss (USD) & $-$2.48 & $-$44.81 & $-$7.06 & $-$10.79\\
\midrule
 0\% Transaction Fee & \citeauthor{gatev_pairs_2006} (\citeyear{gatev_pairs_2006}) & \citeauthor{kim_optimizing_2019} (\citeyear{kim_optimizing_2019}) & RL{$_1$} & RL{$_2$}\\
\midrule
Cumulative Profit & 10.54\% & $-$2.00\% & 9.94\% & 80.92\% \\
Sharpe Ratio & 33.90 & $-$5.76 & 32.74 & 2668.86 \\
Total Action Count & 483 & 43 & 249 & 429\\
Won Action Count & 363 & 23 & 240 & 342 \\
Lost Action Count & 120 & 20 & 9 & 87 \\
Win/Loss Action Ratio & 3.02 & 1.15 & 26.67 & 3.93\\
Max Win Action (USD) & 77.04 & 163.59 & 121.74 & 699.51\\
Max Loss Action (USD) & $-$26.03 & $-$217.54 & $-$21.33 & $-$72.21\\
Avg Win Action Profit/Loss (USD) & 13.07 & 36.69 & 17.77 & 104.25\\
Avg Loss Action Profit/Loss (USD) & $-$2.43 & $-$80.76 & $-$7.06 & $-$16.68\\
\bottomrule
\end{tabular}}
\noindent{\footnotesize{The trading period is from 1 December 2023 to 31 December 2023 with an interest rate of 5.5\%. \acrshort{rl}$_1$ stands for the pair trading that allows \acrlong{rl} to decide upon the investment timing. \acrshort{rl}$_2$ stands for dynamic scaling \acrlong{rl} pair trading that allows the \acrshort{rl} agent to decide both investment timing and quantity. We adopted the \acrshort{ppo} algorithm from \citeauthor{kim_optimizing_2019} (\citeyear{kim_optimizing_2019}) and \acrshort{a2c} for RL$_1$ and RL$_2$.}}
%\end{threeparttable}
\end{table}

Comparison between \acrshort{rl}-based pair trading revealed the relationship between
profitability and actions. Because financial trading is a special case
of the \acrshort{rl} environment, every action in financial trading is
punished by the transaction cost. We notice that profitable
\acrshort{rl} trading often has a lower total trade count and higher profit
per-win trade. That means the \acrshort{rl} is better at spotting
chances to make higher profits. \acrshort{rl}$_2$ pair trading produces
higher profits because of higher average wins from the position
adjustment mechanism. When we adopt the righteous trading method, market volatility and transaction cost play crucial roles in profitable trading. Variable thresholds might not be adaptive to highly volatile markets, and fixed-threshold pair trading could lead to missing trading opportunities. RL with dynamic scaling investment could be a good direction in volatile market conditions if low transaction costs are achievable.

The techniques presented have certain limitations and offer opportunities for future work. One limitation is the relatively limited dataset scope, which could be expanded to include more diverse assets and longer timeframes to improve generalization. Additionally, focusing only on two-leg strategies restricts the potential for complex arbitrage opportunities; expanding to multi-leg strategies would enhance robustness. The computational demand during training can also be resource-intensive, requiring system parameter tuning. The model lacks consideration for transaction costs, which might impact real-world profitability. A lack of direct comparison with traditional models is another shortcoming. Future work could involve developing the Reinforcement Learning (RL) approach to multi-leg strategies, integrating pair formation into the trading process, cross-validating across different environments, and experimenting with alternative reward functions to improve decision-making and risk management. 

\vspace{6pt}

\authorcontributions{Conceptualization, A.M. and H.Y.; methodology, A.M. and H.Y.; software, A.M. and H.Y.; validation, H.Y.; formal analysis, H.Y.; investigation, H.Y.; resources, A.M. and H.Y.; data curation, H.Y.; writing---original draft preparation, H.Y.; writing---review and editing, A.M.; visualization, H.Y.; supervision, A.M.; project administration, A.M.; funding acquisition, Not applicable. All authors have read and agreed to the published version of the manuscript.}

\funding{This research received no external funding
}

\institutionalreview{Not applicable.}

\dataavailability{The original data presented in the study are openly available from Binance Exchange accessed on 8 November 2024 at
 (\url{https://data.binance.vision/}).}

\conflictsofinterest{The authors declare no conflicts of interest.}

\appendixtitles{no}
\appendixstart
\appendix
\section{}
\vspace{-12pt}

\begin{figure}[H]
  \centering
  \begin{subfigure}[b]{0.496\textwidth}
    \includegraphics[width=\textwidth]{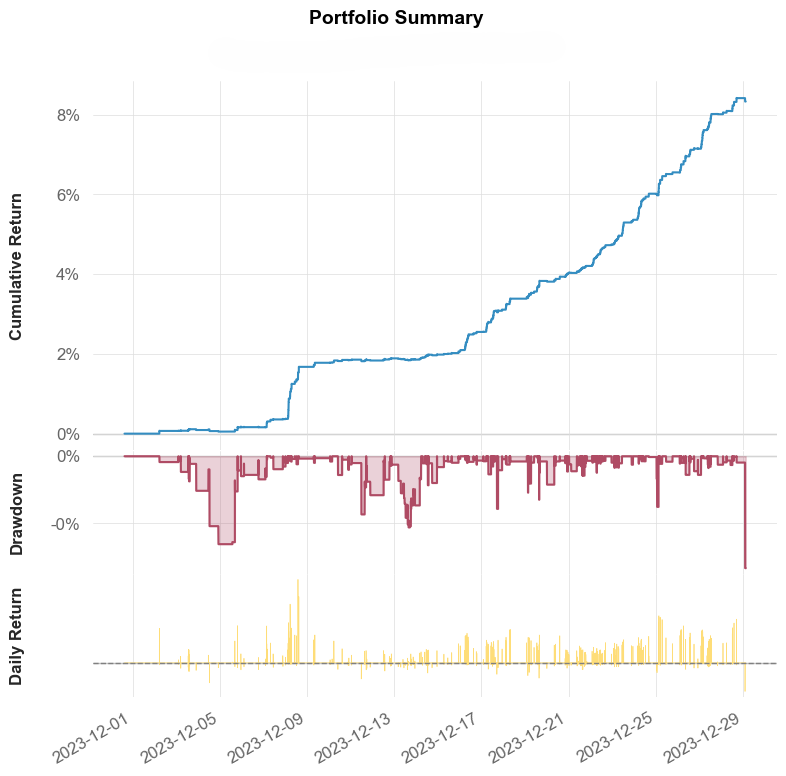}
    \caption{\centering}
    \label{fig:gatevpt}\vspace{+6pt}
  \end{subfigure}
  \hfill
  \begin{subfigure}[b]{0.496\textwidth}
    \includegraphics[width=\textwidth]{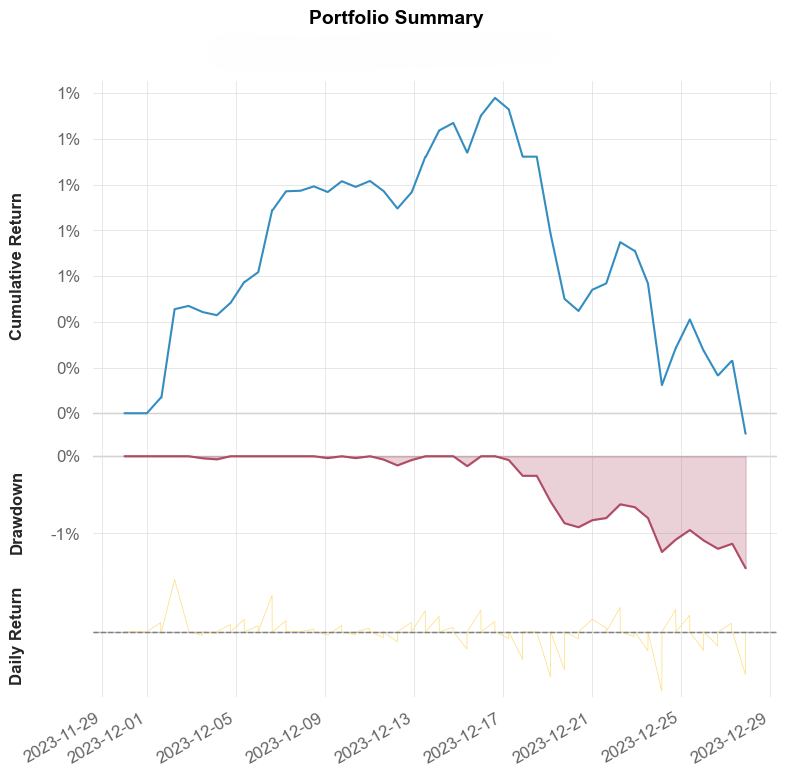}
    \caption{\centering}
    \label{fig:kimpt}\vspace{+6pt}
  \end{subfigure}
  \caption{Comparison of Pair Trading strategies from (\textbf{a}) \citeauthor{gatev_pairs_2006} (\citeyear{gatev_pairs_2006}) and (\textbf{b}) \citeauthor{kim_optimizing_2019} (\citeyear{kim_optimizing_2019}).}
  \label{fig:comparison}
\end{figure}

\begin{adjustwidth}{-\extralength}{0cm}
\printendnotes[custom]
\reftitle{References}
%\centering %% If there is a figure in wide page, please release command \centering

\PublishersNote{}
\end{adjustwidth}


\begin{thebibliography}{999}

\bibitem[\protect\citeauthoryear{AlMahamid and Grolinger}{AlMahamid and
Grolinger}{2021}]{almahamid_reinforcement_2021}
AlMahamid, Fadi, and Katarina Grolinger. 2021.
\newblock Reinforcement learning algorithms: {An} overview and classification.
Paper presented at the 2021 IEEE Canadian Conference on Electrical and Computer Engineering (CCECE), Virtual,  September 12--17;
\newblock pp.\  1--7.

\bibitem[\protect\citeauthoryear{Bellman}{Bellman}{1957}]{bellman_markovian_1957}
Bellman, Richard. 1957.
\newblock A {Markovian} {Decision} {Process}.
\newblock {\em Journal of Mathematics and Mechanics\/}~{6\/}: 679--84. [\href{http://doi.org/10.1512/iumj.1957.6.56038}{CrossRef}]

\bibitem[\protect\citeauthoryear{Bertoluzzo and Corazza}{Bertoluzzo and
Corazza}{2007}]{bertoluzzo_making_2007}
Bertoluzzo, Francesco, and Marco Corazza. 2007.
\newblock Making {Financial} {Trading} by {Recurrent} {Reinforcement}
{Learning}.
\newblock In  {\em
Knowledge-{Based} {Intelligent} {Information} and {Engineering} {Systems}}. Edited by Bruno 
~Apolloni, Robert 
~J. Howlett and Lakhmi 
~Jain.
Lecture {Notes} in {Computer} {Science}. Berlin and Heidelberg: Springer, pp.\  619--26.
\newblock  [\href{http://dx.doi.org/10.1007/978-3-540-74827-4_78}{CrossRef}]

\bibitem[\protect\citeauthoryear{Brogaard, Hendershott, and Riordan}{Brogaard
et~al.}{2014}]{brogaard_high-frequency_2014}
Brogaard, Jonathan, Terrence Hendershott, and Ryan Riordan. 2014.
\newblock High-{Frequency} {Trading} and {Price} {Discovery}.
\newblock {\em The Review of Financial Studies\/}~{27}: 2267--306.
\newblock  [\href{http://dx.doi.org/10.1093/rfs/hhu032}{CrossRef}]

\bibitem[\protect\citeauthoryear{Burgess}{Burgess}{2003}]{burgess_using_2003}
Burgess, A.~Neil. 2003.
\newblock Using {Cointegration} to {Hedge} and {Trade} {International}
{Equities}.
\newblock In {\em Applied {Quantitative} {Methods} for {Trading} and
{Investment}}.  Hoboken: 
John Wiley \& Sons, Ltd., pp.\  41--69.  [\href{http://dx.doi.org/10.1002/0470013265.ch2}{CrossRef}]



\bibitem[\protect\citeauthoryear{Dickey and Fuller}{Dickey and
Fuller}{1979}]{dickey_distribution_1979}
Dickey, David~A., and Wayne~A. Fuller. 1979.
\newblock Distribution of the {Estimators} for {Autoregressive} {Time} {Series}
with a {Unit} {Root}.
\newblock {\em Journal of the American Statistical Association\/}~{74\/}: 427--31.
 [\href{http://dx.doi.org/10.1080/01621459.1979.10482531}{CrossRef}]

\bibitem[\protect\citeauthoryear{Do and Faff}{Do and Faff}{2010}]{do_does_2010}
Do, Binh, and Robert Faff. 2010.
\newblock Does {Simple} {Pairs} {Trading} {Still} {Work}?
\newblock {\em Financial Analysts Journal\/}~{66\/}: 83--95. 
 [\href{http://dx.doi.org/10.2469/faj.v66.n4.1}{CrossRef}]

\bibitem[\protect\citeauthoryear{Dunis and Ho}{Dunis and
Ho}{2005}]{dunis_cointegration_2005}
Dunis, Christian~L., and Richard Ho. 2005.
\newblock Cointegration portfolios of {European} equities for index tracking
and market neutral strategies.
\newblock {\em Journal of Asset Management\/}~{6\/}: 33--52.
\newblock  [\href{http://dx.doi.org/10.1057/palgrave.jam.2240164}{CrossRef}]

\bibitem[\protect\citeauthoryear{Dybvig and Ross}{Dybvig and
Ross}{1989}]{dybvig_arbitrage_1989}
Dybvig, Philip~H., and Stephen~A. Ross. 1989.
\newblock Arbitrage.
\newblock In {\em Finance}.  Edited by John 
~Eatwell, Murray 
~Milgate and Peter 
~Newman.  London: Palgrave Macmillan UK, pp.  57--71.
\newblock  [\href{http://dx.doi.org/10.1007/978-1-349-20213-3_4}{CrossRef}]

\bibitem[\protect\citeauthoryear{Fadok, Boyd, and Warden}{Fadok
et~al.}{1995}]{fadok_air_1995}
Fadok, David~S., John Boyd, and John Warden. 1995.
\newblock Air power’s quest for strategic paralysis.
\newblock  {\em Proceedings of the School of Advanced Airpower Studies\/}. Available online:  \url{https://media.defense.gov/2017/Dec/27/2001861508/-1/-1/0/T_0029_FADOK_BOYD_AND_WARDEN.PDF} (accessed on 8 November 2024)

\bibitem[\protect\citeauthoryear{Gatev, Goetzmann, and Rouwenhorst}{Gatev
et~al.}{2006}]{gatev_pairs_2006}
Gatev, Evan, William~N. Goetzmann, and K.~Geert Rouwenhorst. 2006.
\newblock Pairs {Trading}: {Performance} of a {Relative} {Value} {Arbitrage}
{Rule}. \emph{The~Review~of~Financial~Studies} 19: 797--827. 
\newblock  [\href{http://dx.doi.org/10.1093/rfs/hhj020}{CrossRef}]

\bibitem[\protect\citeauthoryear{Gold}{Gold}{2003}]{gold_fx_2003}
Gold, Carl. 2003.
\newblock {FX} trading via recurrent reinforcement learning.
\newblock {Paper presented at the 2003 IEEE International Conference on Computational Intelligence
for Financial Engineering}, Hong Kong, China, March 20--23,  pp. 363--70,
ISBN 9780780376540.
 [\href{http://dx.doi.org/10.1109/CIFER.2003.1196283}{CrossRef}]

\bibitem[\protect\citeauthoryear{Haarnoja, Zhou, Hartikainen, Tucker, Ha, Tan,
Kumar, Zhu, Gupta, Abbeel, and Levine}{Haarnoja
et~al.}{2019}]{haarnoja_soft_2019}
Haarnoja, Tuomas, Aurick Zhou, Kristian Hartikainen, George Tucker, Sehoon Ha,
Jie Tan, Vikash Kumar, Henry Zhu, Abhishek Gupta, Pieter Abbeel, and et al. 2019.
\newblock Soft {Actor}-{Critic} {Algorithms} and {Applications}. \emph{arXiv}
\newblock arXiv:1812.05905.
 [\href{http://dx.doi.org/10.48550/arXiv.1812.05905}{CrossRef}]

\bibitem[\protect\citeauthoryear{Han, Huang, Xie, Zhang, Lai, and Peng}{Han
et~al.}{2023}]{han_mastering_2023}
Han, Weiguang, Jimin Huang, Qianqian Xie, Boyi Zhang, Yanzhao Lai, and Min
Peng. 2023.
\newblock Mastering {Pair} {Trading} with {Risk}-{Aware} {Recurrent}
{Reinforcement} {Learning}. \emph{arXiv}
\newblock arXiv:2304.00364.

\bibitem[\protect\citeauthoryear{Huang}{Huang}{2018}]{huang_financial_2018}
Huang, Chien~Yi. 2018.
\newblock Financial {Trading} as a {Game}: {A} {Deep} {Reinforcement}
{Learning} {Approach}. \emph{arXiv}
\newblock arXiv:1807.02787.
 [\href{http://dx.doi.org/10.48550/arXiv.1807.02787}{CrossRef}]

\bibitem[\protect\citeauthoryear{Huck}{Huck}{2010}]{huck_pairs_2010}
Huck, Nicolas. 2010.
\newblock Pairs trading and outranking: {The} multi-step-ahead forecasting
case.
\newblock {\em European Journal of Operational Research\/}~{207\/}:
1702--16.
\newblock  [\href{http://dx.doi.org/10.1016/j.ejor.2010.06.043}{CrossRef}]

\bibitem[\protect\citeauthoryear{Kim and Kim}{Kim and
Kim}{2019}]{kim_optimizing_2019}
Kim, Taewook, and Ha~Young Kim. 2019.
\newblock Optimizing the {Pairs}-{Trading} {Strategy} {Using} {Deep}
{Reinforcement} {Learning} with {Trading} and {Stop}-{Loss} {Boundaries}.
\newblock {\em Complexity\/}~{2019}: e3582516.
 [\href{http://dx.doi.org/10.1155/2019/3582516}{CrossRef}]









\bibitem[\protect\citeauthoryear{Liu, Yang, Gao, and Wang}{Liu
et~al.}{2021}]{liu_finrl_2021}
Liu, Xiao-Yang, Hongyang Yang, Jiechao Gao, and Christina~Dan Wang. 2021.
\newblock {FinRL}: {Deep} {Reinforcement} {Learning} {Framework} to {Automate}
{Trading} in {Quantitative} {Finance}.
\newblock {Paper presented at the Proceedings of the {Second} {ACM} {International}
{Conference} on {AI} in {Finance}}, Virtual Event, November 3--5, pp.\  1--9.
\newblock  
 [\href{http://dx.doi.org/10.1145/3490354.3494366}{CrossRef}]

\bibitem[\protect\citeauthoryear{Liu, Yang, Chen, Zhang, Yang, Xiao, and
Wang}{Liu et~al.}{2022a}]{liu_finrl_2022}
Liu, Xiao-Yang, Hongyang Yang, Qian Chen, Runjia Zhang, Liuqing Yang, Bowen
Xiao, and Christina~Dan Wang. 2022a.
\newblock {FinRL}: {A} {Deep} {Reinforcement} {Learning} {Library} for
{Automated} {Stock} {Trading} in {Quantitative} {Finance}. \emph{arXiv}
\newblock arXiv:2011.09607.
 [\href{http://dx.doi.org/10.48550/arXiv.2011.09607}{CrossRef}]
 
\bibitem[\protect\citeauthoryear{Liu, Xia, Rui, Gao, Yang, Zhu, Wang, Wang, and
Guo}{Liu et~al.}{2022b}]{liu_finrl-meta_2022}
Liu, Xiao-Yang, Ziyi Xia, Jingyang Rui, Jiechao Gao, Hongyang Yang, Ming Zhu,
Christina~Dan Wang, Zhaoran Wang, and Jian Guo. 2022b.
\newblock {FinRL}-{Meta}: {Market} {Environments} and {Benchmarks} for
{Data}-{Driven} {Financial} {Reinforcement} {Learning}. \emph{arXiv}
\newblock arXiv:2211.03107.
 [\href{http://dx.doi.org/10.2139/ssrn.4253139}{CrossRef}]
 
 
\bibitem[\protect\citeauthoryear{Lucarelli and Borrotti}{Lucarelli and
Borrotti}{2019}]{lucarelli_deep_2019}
Lucarelli, Giorgio, and Matteo Borrotti. 2019.
\newblock A {Deep} {Reinforcement} {Learning} {Approach} for {Automated}
{Cryptocurrency} {Trading}.
\newblock In {\em Artificial {Intelligence} {Applications} and {Innovations}}.
{IFIP} {Advances} in {Information} and {Communication} {Technology}.  Edited by John 
~MacIntyre, Ilias 
~Maglogiannis, Lazaros 
~Iliadis and Elias 
~Pimenidis. Cham: Springer International Publishing,
pp.\  247--58.
\newblock  [\href{http://dx.doi.org/10.1007/978-3-030-19823-7_20}{CrossRef}]

\bibitem[\protect\citeauthoryear{Mandelbrot}{Mandelbrot}{1967}]{mandelbrot_variation_1967}
Mandelbrot, Benoit. 1967.
\newblock The {Variation} of {Some} {Other} {Speculative} {Prices}.
\newblock {\em The Journal of Business\/}~{40\/}: 393--413. [\href{http://dx.doi.org/10.1086/295006}{CrossRef}]


\bibitem[\protect\citeauthoryear{Maringer and Ramtohul}{Maringer and
Ramtohul}{2012}]{maringer_regime-switching_2012}
Maringer, Dietmar, and Tikesh Ramtohul. 2012.
\newblock Regime-switching recurrent reinforcement learning for investment
decision making.
\newblock {\em Computational Management Science\/}~{9\/}: 89--107.
\newblock  [\href{http://dx.doi.org/10.1007/s10287-011-0131-1}{CrossRef}]

\bibitem[\protect\citeauthoryear{Meng and Khushi}{Meng and
Khushi}{2019}]{meng_reinforcement_2019}
Meng, Terry~Lingze, and Matloob Khushi. 2019.
\newblock Reinforcement {Learning} in {Financial} {Markets}.
\newblock {\em Data\/}~{4\/}: 110.
 [\href{http://dx.doi.org/10.3390/data4030110}{CrossRef}]

\bibitem[\protect\citeauthoryear{Mnih, Kavukcuoglu, Silver, Graves, Antonoglou,
Wierstra, and Riedmiller}{Mnih et~al.}{2013}]{mnih_playing_2013}
Mnih, Volodymyr, Koray Kavukcuoglu, David Silver, Alex Graves, Ioannis
Antonoglou, Daan Wierstra, and Martin Riedmiller. 2013.
\newblock Playing {Atari} with {Deep} {Reinforcement} {Learning}. \emph{arXiv}
\newblock arXiv:1312.5602.
 [\href{http://dx.doi.org/10.48550/arXiv.1312.5602}{CrossRef}]

\bibitem[\protect\citeauthoryear{Mohammadshafie, Mirzaeinia, Jumakhan, and
Mirzaeinia}{Mohammadshafie et~al.}{2024}]{mohammadshafie_deep_2024}
Mohammadshafie, Alireza, Akram Mirzaeinia, Haseebullah Jumakhan, and Amir
Mirzaeinia. 2024.
\newblock Deep {Reinforcement} {Learning} {Strategies} in {Finance}: {Insights}
into {Asset} {Holding}, {Trading} {Behavior}, and {Purchase} {Diversity}. \emph{arXiv}
\newblock arXiv:2407.09557.  [\href{http://dx.doi.org/10.48550/arXiv.2407.09557}{CrossRef}]

\bibitem[\protect\citeauthoryear{Perlin}{Perlin}{2007}]{perlin_m_2007}
Perlin, Marcelo. 2007.
\newblock M of a {Kind}: {A} {Multivariate} {Approach} at {Pairs} {Trading}.
Available online: {\url{https://doi.org/10.2139/ssrn.952782}} (accessed on 8 November 2024).

\bibitem[\protect\citeauthoryear{Perlin}{Perlin}{2009}]{perlin_evaluation_2009}
Perlin, Marcelo~Scherer. 2009.
\newblock Evaluation of pairs-trading strategy at the {Brazilian} financial
market.
\newblock {\em Journal of Derivatives \& Hedge Funds\/}~{15\/}:
122--36.
\newblock  [\href{http://dx.doi.org/10.1057/jdhf.2009.4}{CrossRef}]

\bibitem[\protect\citeauthoryear{Pricope}{Pricope}{2021}]{pricope_deep_2021}
Pricope, Tidor-Vlad. 2021.
\newblock Deep {Reinforcement} {Learning} in {Quantitative} {Algorithmic}
{Trading}: {A} {Review}. \emph{arXiv}
\newblock arXiv:2106.00123.
 [\href{http://dx.doi.org/10.48550/arXiv.2106.00123}{CrossRef}]

\bibitem[\protect\citeauthoryear{Raffin, Hill, Gleave, Kanervisto, Ernestus,
and Dormann}{Raffin et~al.}{2021}]{raffin_stable-baselines3_2021}
Raffin, Antonin, Ashley Hill, Adam Gleave, Anssi Kanervisto, Maximilian
Ernestus, and Noah Dormann. 2021.
\newblock Stable-baselines3: {Reliable} reinforcement learning implementations.
\newblock {\em Journal of Machine Learning Research\/}~{22\/}: 1--8.

\bibitem[\protect\citeauthoryear{Sarmento and Horta}{Sarmento and
Horta}{2020}]{sarmento_enhancing_2020}
Sarmento, Simão~Moraes, and Nuno Horta. 2020.
\newblock Enhancing a {Pairs} {Trading} strategy with the application of
{Machine} {Learning}.
\newblock {\em Expert Systems with Applications\/}~{158}: 113490.
\newblock  [\href{http://dx.doi.org/10.1016/j.eswa.2020.113490}{CrossRef}]

\bibitem[\protect\citeauthoryear{Schulman, Wolski, Dhariwal, Radford, and
Klimov}{Schulman et~al.}{2017}]{schulman_proximal_2017}
Schulman, John, Filip Wolski, Prafulla Dhariwal, Alec Radford, and Oleg Klimov.
2017.
\newblock Proximal {Policy} {Optimization} {Algorithms}. \emph{arXiv}
\newblock arXiv:1707.06347.
 [\href{http://dx.doi.org/10.48550/arXiv.1707.06347}{CrossRef}]

\bibitem[\protect\citeauthoryear{Sharpe}{Sharpe}{1964}]{sharpe_capital_1964}
Sharpe, William~F. 1964.
\newblock Capital {Asset} {Prices}: {A} {Theory} of {Market} {Equilibrium}
{Under} {Conditions} of {Risk}.
\newblock {\em The Journal of Finance\/}~{19\/}: 425--42.
 [\href{http://dx.doi.org/10.1111/j.1540-6261.1964.tb02865.x}{CrossRef}]


\bibitem[\protect\citeauthoryear{{David Silver, Demis Hassabis}}{{Silver and
Hassabis}}{2016}]{david_silver_demis_hassabis_alphago_2016}
{Silver, David, and Demis Hassabis}. 2016.
\newblock {AlphaGo}: {Mastering} the ancient game of {Go} with {Machine}
{Learning}.  Available online: {\url{https://research.google/blog/alphago-mastering-the-ancient-game-of-go-with-machine-learning/}} (accessed on 8 November 2024). 

\bibitem[\protect\citeauthoryear{Sutton and Barto}{Sutton and
Barto}{2018}]{sutton_reinforcement_2018}
Sutton, Richard~S., and Andrew~G. Barto. 2018.
\newblock {\em Reinforcement Learning: {An} Introduction}.
\newblock Cambridge: MIT Press.

\bibitem[\protect\citeauthoryear{Vergara and Kristjanpoller}{Vergara and
Kristjanpoller}{2024}]{vergara_deep_2024}
Vergara, Gabriel, and Werner Kristjanpoller. 2024.
\newblock Deep reinforcement learning applied to statistical arbitrage
investment strategy on cryptomarket.
\newblock {\em Applied Soft Computing\/}~{153}: 111255.
\newblock  [\href{http://dx.doi.org/10.1016/j.asoc.2024.111255}{CrossRef}]

\bibitem[\protect\citeauthoryear{Wang, Sandås, and Beling}{Wang
et~al.}{2021}]{wang_improving_2021}
Wang, Cheng, Patrik Sandås, and Peter Beling. 2021.
\newblock Improving {Pairs} {Trading} {Strategies} via {Reinforcement}
{Learning}.
\newblock Paper Presented at the {2021 {International} {Conference} on {Applied} {Artificial}
{Intelligence} ({ICAPAI})}, Halden, Norway, May 19--21, pp.\  1--7.
\newblock  [\href{http://dx.doi.org/10.1109/ICAPAI49758.2021.9462067}{CrossRef}]

\bibitem[\protect\citeauthoryear{Yang and Malik}{Yang and
Malik}{2024}]{yang_optimal_2024}
Yang, Hongshen, and Avinash Malik. 2024.
\newblock Optimal market-neutral currency trading on the cryptocurrency
platform. \emph{arXiv}
\newblock arXiv:2405.15461.
 [\href{http://dx.doi.org/10.48550/arXiv.2405.15461}{CrossRef}]

\bibitem[\protect\citeauthoryear{Zhang and Maringer}{Zhang and
Maringer}{2016}]{zhang_using_2016}
Zhang, Jin, and Dietmar Maringer. 2016.
\newblock Using a {Genetic} {Algorithm} to {Improve} {Recurrent}
{Reinforcement} {Learning} for {Equity} {Trading}.
\newblock {\em Computational Economics\/}~{47\/}: 551--67.
\newblock  [\href{http://dx.doi.org/10.1007/s10614-015-9490-y}{CrossRef}]

\bibitem[\protect\citeauthoryear{Zhang, Zohren, and Stephen}{Zhang
et~al.}{2020}]{zhang_deep_2020}
Zhang, Zihao, Stefan Zohren, and Roberts Stephen. 2020.
\newblock Deep {Reinforcement} {Learning} for {Trading}.
\newblock {\em The Journal of Financial Data Science\/} 2: 25--40. 
 [\href{http://dx.doi.org/10.3905/jfds.2020.1.030}{CrossRef}]


\end{thebibliography}
\end{document}